\documentclass[prb,twocolumn,showpacs,amsmath,amssymb,floatfix]{revtex4}
\usepackage{graphicx}
\usepackage{dcolumn} 
\usepackage{bm}
\usepackage{amsmath}
\usepackage{latexsym}
\usepackage{amsfonts}
\usepackage{amssymb}
\usepackage[rightcaption]{sidecap}
\usepackage{subfigure}
\DeclareGraphicsExtensions{.pdf,.gif,.jpg}
\newcommand{\be}{\begin{equation}}
\newcommand{\ee}{\end{equation}}
\newcommand{\bea}{\begin{eqnarray}}
\newcommand{\eea}{\end{eqnarray}}
\newcommand{\nn} {\nonumber}
\renewcommand{\vr} {{\bf r}}

\newcommand{\Tr}{ {\rm Tr} \, }

\def\G{\Gamma}

\def\ve{\varepsilon}

\def\l{\lambda}

\def\S{\Sigma}

\def\vf{\varphi}

\def\w{\omega}

\def\bra{\langle}
\def\ket{\rangle}

\def\xc{{\rm xc}}
\def\x{{\rm x}}

\def\Tr{{\rm Tr}\,}

\begin{document}
\title{Exact-exchange kernel of time-dependent density functional 
theory: Frequency dependence and photoabsorption spectra of atoms}
\date{\today}
\author{Maria Hellgren}
\author{Ulf von Barth}
\affiliation{Mathematical Physics, Institute of Physics, Lund 
University, S\"olvegatan 14A, S-22362 Lund, Sweden}  
\date{\today}              
\begin{abstract}
In this work we have calculated excitation energies and 
photoionization cross sections of Be and Ne in the exact-exchange (EXX) approximation 
of time-dependent density functional theory (TDDFT). 
The main focus has been on the frequency dependence of the EXX kernel 
and on how it affects the spectrum as compared to the corresponding 
adiabatic approximation. We show that for some discrete excitation 
energies the frequency dependence is essential to reproduce the 
results of time-dependent Hartree-Fock theory. Unfortunately, we have found 
that the EXX approximation breaks down completely at higher energies, producing a
response function with the wrong analytic structure and making inner-shell excitations 
disappear from the calculated spectra. We have traced this 
failure to the existence of vanishing eigenvalues of the Kohn-Sham non-interacting
response function. Based on the adiabatic TDDFT formalism we propose a new way 
of deriving the Fano parameters of autoionizing resonances. 
\end{abstract}
\maketitle
\section{Introduction}
The photoabsorption spectrum of an atom or molecule reveals detailed
information about the structure and dynamics of its constituent 
electrons. Important phenomena, such as Fano resonances\cite{fano} and 
multiple-particle excitations, which can solely be attributed to the electron-
electron interaction, has been under extensive experimental and theoretical 
investigation for several decades. To describe and quantify these exciting 
many-body effects is, however, still a major theoretical challenge. Traditional 
approaches to this problem are based on infinite-order many-body perturbation 
expansions which lead to complicated integral equations already at the Hartree-Fock 
level of
approximation. This level of approximation is also referred to as the 
random phase approximation 
with exchange (RPAE) and was extensively used by 
Amusia\cite{amusiaulf,amusiacase} 
and by Wendin\cite{wendin} for the description 
of atomic photoabsorption spectra. From a mathematical or numerical 
point of view this entails the solution of an integral equation known 
as the Bethe-Salpeter equation for a four-point vertex function, the 
solution of which is computationally very demanding in low-symmetry systems 
like larger molecules or nano structures. 

Since the pioneering work of Ando\cite{ando} and Zangvill 
and Soven\cite{zs} in the late seventies,
time-dependent density functional theory\cite{peuk,rg0} (TDDFT) has emerged as a 
more convenient route
toward the calculation of spectral properties. Within TDDFT, as it is 
presently
formulated, only two-point correlation functions appear and, in 
addition,
one arrives at explicit expressions for the linear density response
function, thus avoiding the need for integral equations. The price to
pay is the difficulties associated with obtaining approximations to 
the frequency-dependent exchange-correlation (XC) kernel, constituting 
one of the basic elements of TDDFT. In most calculations until today one uses a 
static, frequency-independent, kernel. These so called adiabatic 
approximations have, in many cases, produced relatively accurate and 
important results - especially for static
properties or small excitation energies. There are, however, many 
cases
where the inadequacies of the adiabatic approximations are easily
discernible. 

Until recently, no systematic way of obtaining successively
better approximations to the XC kernel of TDDFT has been available. 
By means of the 
variational formulation of many-body theory\cite{abl} we have 
recently developed a tool 
for constructing approximations to the XC kernel, which not only 
incorporates the important energy dependence of the kernel but does it in a way so as 
to guarantee the fulfillment of many sum rules and conservation 
laws.\cite{tddftvar} As discussed in these previous publications the 
variational approach is not uniquely defined and some 
implementations are preferable to others. For instance, starting from the functional of 
Luttinger and Ward\cite{lw} (LW) is usually a better approach as compared to starting 
from the Klein functional.\cite{klein}
The simplest approximate kernel derived from the latter functional 
is that of the time-dependent exact-exchange (TDEXX) approximation. 
This approximation can be considered as an attempt to mimic the time-dependent 
Hartree-Fock approximation within a TDDFT framework, a point which has been discussed 
at length in previous work.\cite{kvB,hvb2} The exact-exchange (EXX) kernel has, however, 
yet received little attention especially with regard to its ability to describe spectral 
properties which necessitates a calculation on the real frequency axis. Previous investigations have either addressed the adiabatic limit,\cite{higb2002,pgg}  focused on extended systems at low frequencies\cite{kg2} or most recently carried out model calculations in the time domain.\cite{Wiull2008}

The present paper presents a numerical and analytical investigation of the fully 
frequency-dependent EXX kernel of closed-shell atoms. It is the fourth and 
latest in a series of papers by us 
concerned with approximations obtained from the variational scheme to 
TDDFT. In the first
paper\cite{kvB} we calculated the EXX kernel of the electron
gas. From that kernel we obtained the interacting density response
function of the gas, a quantity which enabled us to calculate the
correlation energy via an integration over the strength of the Coulomb
interaction. The results turned out to be rather close to accurate Monte-Carlo 
results. From the response function
one can also indirectly obtain a local vertex function which can be
used to calculate the electron self-energy and thus the one-electron
Green function. Unfortunately, this local vertex turned out to be too
simple minded to remedy the infamous band-narrowing 
problem\cite{whaplu88,mahser89} in simpler metals - and in transition 
metals for that matter.\cite{ferdiulf} In the 
second paper\cite{hvB} we went beyond the EXX approximation and calculated 
the XC potential at the level of the self-consistent GW
approximation in atoms. This potential turned out to be superior to 
that of the second-order M\o ller-Plesset (MP2) 
approximation,\cite{engeljiang} at least in the
outskirts of the atoms. The resulting ionization potentials turned out
to be surprisingly accurate. In the third paper\cite{hvb2} we 
calculated the 
EXX kernel at imaginary frequencies and used it to calculate static
polarizabilities, van der-Waals coefficient and correlation energies 
of closed-shell atoms. The polarizabilities and van der-Waals 
coefficients were very similar to those of time-dependent Hartree-Fock theory and
thus not very accurate. (See also Ref. \onlinecite{shh2006}.) On the other hand, 
we obtained better than 95\% of the correlation energies which should be 
compared to errors of the order of a factor of two within the RPA. In contrast, 
our calculated discrete particle-conserving excitation energies were not far from
those of the RPA. We attributed this partial failure to a poor
description of the ground state within the EXX approximation.

It is interesting to investigate the performance of the TDEXX approximation 
as far as concerns the description of photoabsorption spectra. In previous papers 
we were, relying on a new numerical technique based on cubic splines. 
While this technique turned out to be ideally suited for the necessary 
inversion of the non-interacting density response function and also for the 
calculation of discrete excitation energies, the method was less 
effective in the case of continuous spectra. This deficiency is mainly 
a result of a too sparse and unpredictable sampling of the continuum 
especially at larger energies, something which would require a 
prohibitively large number of splines to remedy. 
For these reasons we have in the present work designed interpolation schemes 
and new numerical methods for directly extracting the important parameters 
determining the shape of the Fano resonances\cite{fano} 
resulting from the existence of quasi discrete inner-shell levels interacting 
with  continua. 

While we have seen that the energy dependence of the EXX kernel can improve 
on many of the deficiencies of typical adiabatic (read
static) approximations we have in the present work made a very
important although somewhat disappointing discovery. The TDEXX 
completely fails to describe parts of the spectra, especially at higher energies.
At least if the spectra are calculated from Fourier transforms to
frequency space as most researches consider to be the most natural
approach.\cite{grosskohn} Moreover, this failure is not connected, e.g., to the 
neglect of processes describing correlation effects. The failure is caused by
the existence of zero eigenvalues of the non-interacting response
function at certain energies. Any method for constructing an XC kernel
starting from the linearized Sham-Schl{\"u}ter (LSS) equation will suffer
from the same ailment no matter how much of correlation one is trying 
to account for. And those variational approaches to many-body theory 
which build on the Klein functional will always result in the LSS 
equation, suggesting that a reasonable theory for constructing better 
kernels should start from a more elaborate variational functional 
like, e.g., that due to Luttinger and Ward. But this is, of course, 
only speculation at the present stage. Such investigations will be the 
topics of future works.

The paper is organized as follows. In Sec. II we present the linear response 
matrix equation for excitation energies and oscillator strengths and discuss 
how a frequency-dependent kernel might influence the results. In Sec. III we present 
the EXX equations. We also discuss possible generalizations and point out inherent 
limitations and potential problems. In Sec. IV the numerical method is briefly discussed. 
Sec. V is devoted to numerical results with a presentation of the EXX kernel
and the consequences of zero eigenvalues in the KS response function. The 
excitation energies and continuum spectra of Be and Ne are also presented and 
compared to the results of the RPA and the adiabatic EXX (AEXX) approximations. 
In Sec. VI we give our conclusion and summarize our findings and, finally, in the 
Appendix we derive the Fano-profile formula\cite{fano} using the adiabatic 
TDDFT formalism. 
\section{Linear response equation}
\label{lre}
Within TDDFT the electronic linear density response function $\chi$ 
is given by 
\be
\chi=\chi_s+\chi_s(v+f_\xc)\chi,
\label{rpafxc}
\ee
where $\chi_s$ is the Kohn-Sham (KS) linear density response 
function, $v$ is the Coulomb interaction 
and $f_\xc$ is the XC kernel defined as the functional derivative of 
the XC potential $v_\xc$,
\be
f_\xc=\frac{\delta v_\xc}{\delta n}.
\ee
Eq. (\ref{rpafxc}) can be considered as a matrix equation in $\vr$ and $\vr'$ or in any other suitable basis.

The KS linear density response function or, equivalently, the 
retarded KS polarization propagator,  
can be written as
\be
\chi_s(\vr,\vr',\w)=\sum_{q}\frac{\tilde{f}_q(\vr)\tilde{f}^{*}_{q}(\vr')}{(\w+i0^{+})^2-\w_q^2}.
\ee
Here, $q=(k,\mu)$ is a composite index, in which $k$ labels an 
occupied state and $\mu$ an unoccupied state. The excitation function 
$f_q=\vf^*_\mu(\vr)\vf_k(\vr)$ is a 
product of an occupied and an unoccupied KS orbital and 
$\w_q=\epsilon_\mu-\epsilon_k$ is a KS excitation energy. The tilde in 
$\tilde{f}_q$ signifies a multiplication by $2\sqrt{\w_q}$. With 
$\chi_s$ in this representation, Eq. (\ref{rpafxc}) for the full 
$\chi$ can be rewritten as\cite{tddftbook}
\be
\chi(\vr,\vr',\w)=\sum_{qq'}\tilde{f}_q(\vr)[(\w+i0^{+})^2\mathbf{I}-\mathbf{V}]^{-1}_{qq'}\tilde{f}^{*}_{q'}(\vr')
\label{m2}
\ee
where $\mathbf{I}$ is the identity matrix and 
\bea
V_{qq'}=\w^2_{q}\delta_{qq'}+\bra \tilde{f}_q|v+f_{\xc}(\w)|\tilde{f}_{q'}\ket.
\label{vmatris}
\eea 
The second term represents an integral over all spatial variables of 
$v+f_\xc$ multiplied with two excitation functions $\tilde{f}_q$ and 
$\tilde{f}^{*}_{q'}$. The matrix $\mathbf{V}$ is symmetric and if, in 
addition, $f_{\rm xc}$ is frequency-independent and hence real, 
$\mathbf{V}$ can be diagonalized as 
$\mathbf{U^{\dag}}\mathbf{V}\mathbf{U}=\mathbf{Z}^2$, where 
$\mathbf{Z}$ is a diagonal matrix. Then Eq. (\ref{m2}) simplifies to
\be
\chi(\vr,\vr',\w)=\sum_{q}\frac{\tilde{g}_q(\vr)\tilde{g}^{*}_{q}(\vr')}{(\w+i0^{+})^2-Z^2_q}.
\label{chi}
\ee
Here, $\tilde{g}_q(\vr)=\sum_{q'}U_{qq'}\tilde{f}_{q'}(\vr)$, where 
$\mathbf{U}$ is the unitary matrix which diagonalizes $\mathbf{V}$. 
The square root of the eigenvalues can be interpreted as the new 
transition frequencies and $\tilde{g}_q$ as the excitation 
amplitudes.\cite{casbook}   
For most known, or generally used, adiabatic kernels the matrix  
$\mathbf{V}$ is dominated by the Coulomb interaction and therefore 
its eigenvalues are positive. As a consequence $\chi$ has the correct 
analytic structure with a positive spectral function. It is also easy 
to see that both $\chi$ and $\chi_s$ have the same large frequency 
behavior and thus both obey the $f$-sum rule.  

A frequency-dependent $f_{\xc}$ with the correct analytic structure 
requires $f_{\xc}$ to have both real and imaginary parts. The matrix 
$\mathbf{V}$ is then not Hermitian. It is, however, still symmetric 
and can, in general, be diagonalized according to $\mathbf{U^{\rm 
T}}\mathbf{V}\mathbf{U}=\mathbf{Z}^2$, where the eigenvalues and 
eigenvectors now can be both complex and $\w$-dependent.  A 
particular consequence of having a frequency-dependent kernel is that 
more zeros in the denominator of Eq. (\ref{chi}) can be generated, 
which would imply more resonances in $\chi$. These resonances could 
correspond to multiple-particle excitations or other excitations of 
collective nature. Unfortunately, an approximate frequency dependence 
is not guaranteed to yield a positive spectral function. It could 
also lead to 
poles in the upper half of the complex plane, thus destroying the 
analytic structure of $\chi$. We will later demonstrate that this is 
indeed a reality to be considered.
\section{Exact-exchange equations}
The derivation of the EXX potential and kernel has been given in 
several previous publications by us\cite{tddftvar,hvB,hvb2} and 
others.\cite{casida,pgg} Our choice of approach is to start from that 
variational approach to many-body theory which builds on the
Klein functional.\cite{klein} This functional of the Green function gives us the 
total action of the system and it contains the functional $\Phi$ 
whose derivative with respect to the Green function gives the 
electronic self energy, $\S=\frac{\delta\Phi}{\delta G}$.\cite{baym} When 
possible Green functions are restricted to those which can be 
generated by local multiplicative potentials the action becomes a 
functional of the density. The stationary property of the action with 
respect to variations in the particle density yields the linearized Sham-Schl\"uter 
(LSS) equation\cite{ss} for the KS potential $v_{\xc}$ (with $\vr_1 t_1\rightarrow 1$):
\be
\int \chi_s(1,2)v_{\xc}(2)d2=\int \S_s(2,3)\Lambda(3,2;1)d2d3,
\label{lsseq}
\ee
where $\S_s$ is the self-energy calculated with KS 
orbitals generated by $v_\xc$ and 
$$
i\Lambda(3,2;1)=\frac{\delta G_s(3,2)}{\delta V(1)}=G_s(3,1)G_s(1,2).
$$
Here, $G_s$ is the KS Green function and $V$ is the total effective 
potential of the KS system. 
A further variation of the LSS equation with respect to the potential $V$ 
results in an equation 
for $f_\xc$:
\begin{eqnarray}
&&\int \chi_s(1,2)f_{\xc}(2,3)\chi_s(3,4)d2d3\nn\\
&&\,\,\,\,\,\,\,\,\,\,=\int \frac{\delta\S_s(2,3)}{\delta 
V(4)}\Lambda(3,2;1)d2d3\nn\\
&&\,\,\,\,\,\,\,\,\,\,\,\,\,\,+\int 
\Lambda(1,2;4)\Delta(2,3)G_s(3,1)d2d3\nn\\
&&\,\,\,\,\,\,\,\,\,\,\,\,\,\,+\int 
G_s(1,2)\Delta(2,3)\Lambda(3,1;4)d2d3,
\label{fxceq}
\end{eqnarray}
where $\Delta(2,3)=\S_s(2,3)-v_{{\xc}}(2)\delta(2,3)$. 

\begin{figure}
\includegraphics[width=8.5cm, clip=true]{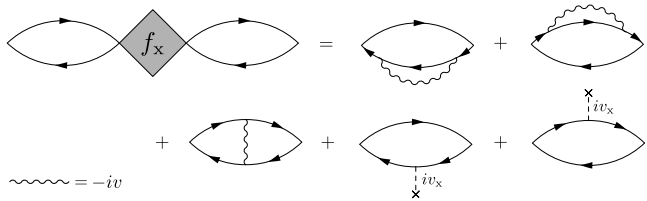}\\
\caption{Diagrammatic representation of the equation for  $f_{\rm 
x}$. }
\label{diagram}
\end{figure}

In the EXX approximation one chooses the HF approximation for the functional $\Phi$,
\be
\Phi=\frac{i}{2}\Tr\left[GGv\right],
\ee 
and the variation of the resulting self energy with respect to $V$
becomes
$$
\frac{\delta \S^{\rm x}_s(2,3)}{\delta V(4)}=-v(2,3)\Lambda(2,3;4).
$$
The terms on the right hand side of Eq. (\ref{fxceq}) are, in this 
approximation, represented diagrammatically in Fig.~\ref{diagram}.
The first order vertex diagram, the first on the second row in Fig.~\ref{diagram}), which will be referred to as $R_V$, 
is given in terms of KS orbitals $\vf_{k}$ and eigenvalues $\ve_k$ as
\begin{subequations}
\bea
&&R_V(\vr,\vr',z)=-2\sum_{k_1k_2}\sum_{k'_1k'_2}\vf_{k_1}(\vr)\vf_{k_2}^*(\vr)\vf_{k'_1}^*(\vr')\vf_{k'_2}(\vr')\nonumber\\
&&\,\,\,\,\,\,\,\,\times\bra 
k_1k'_2|v|k'_1k_2\ket\frac{(n_{k_1}-n_{k_2})(n_{k'_1}-n_{k'_2})}{(z+\ve_{k_2}-\ve_{k_1})(z+\ve_{k'_2}-\ve_{k'_1})}
\eea
where $z=\w+i0^+$ and $n_k$ is the occupation number of state $k$.  
The other four diagrams are self-energy terms and their sum, referred 
to as $R_\S$, is:
\bea
&&R_\S(\vr,\vr',z)=\!\!\!\sum_{k_1k_2k_3}\bra 
k_2|\Delta|k_3\ket\vf_{k_1}(\vr)\vf_{k_1}^*(\vr')\vf_{k_2}(\vr)\vf_{k_3}^*(\vr')\nonumber\\
&&\,\,\,\,\,\,\,\,\,\,\,\,\,\,\,\,\,\,\,\,\,\,\,\,\,\,\,\,\,\,\,\,\,\,\,\,\,\,\,\,\,\,\,\,\times\frac{4}{\ve_{k_3}-\ve_{k_2}}\left\{\frac{(n_{k_3}-n_{k_1})(\ve_{k_1}-\ve_{k_3})}{z^2-(\ve_{k_1}-\ve_{k_3})^2}\right.\nonumber\\
&&\,\,\,\,\,\,\,\,\,\,\,\,\,\,\,\,\,\,\,\,\,\,\,\,\,\,\,\,\,\,\,\,\,\,\,\,\,\,\,\,\,\,\,\,\left.-\frac{(n_{k_2}-n_{k_1})(\ve_{k_1}-\ve_{k_2})}{z^2-(\ve_{k_1}-\ve_{k_2})^2}\right\}
\eea
\label{fxekv}
\end{subequations}

We see immediately that the poles of $R_V$ and $R_\S$ are located at 
the KS eigenvalue differences, just as in the case of $\chi_s$. The 
EXX approximation can, therefore, not describe multiple-particle 
excitations. In order to incorporate those one has to include 
diagrams which contain new poles, as obtained, e.g., by choosing $\S$ 
in the GW approximation. The new poles are in that approximation 
generated by the dynamically screened interaction, $W$.

It is apparent from Eq. (\ref{rpafxc}) that only the quantity 
$\chi_sf_\xc$ is needed to obtain the spectrum. Consequently, it is 
sufficient to invert $\chi_s$ once in Eq. (\ref{fxceq}). It has, 
however, been known for some time that $\chi_s$ is not always 
invertible due to the existence of nontrivial vanishing eigenvalues 
at particular frequencies.\cite{mearns} At these frequencies there 
exists an external perturbation which does not produce a density 
response to first order. Thus, in the unlikely case that $R=R_V+R_\S $ 
has zero eigenvalues at the same frequencies the kernel $f_\x$ diverges.\cite{fot1} 
This can have a drastic effect on the calculated spectra as we will see later.
\section{Numerical approach}
For the numerical implementation we have used an approach based on cubic 
splines as radial basis functions. This approach was 
used also in previous works\cite{hvB,hvb2} and showed to be ideally suited 
for solving equations where an inversion of $\chi_{s}$ is needed. 
In short, a cubic spline is a piecewise third order polynomial with compact support. 
Defined on four sub-intervals, and hence composed of four different third order 
polynomials with in total 16 unknown constants, it is uniquely determined up to a 
multiplicative constant by imposing continuity up to the second derivative. 
A mesh is distributed on the $r$-axis up to a finite $r_{\rm max}$ 
and the basis set is formed by constructing a spline starting at every mesh 
point and extending over four intervals. In this way the splines only overlap with 
its three nearest neighbors on either side. The basis set is complete on the interval 
$\left[0,r_{\rm max}\right]$ in the limit $N\rightarrow \infty$, where $N$ is 
the number of splines. A review on the use of general B-splines in electronic 
structure calculations can be found in Ref. \onlinecite{splines}. 

The numerical solution of Eqs. (\ref{lsseq}-\ref{fxceq}) involves the calculation
of products of KS orbitals. The completeness of the spline basis allows us to 
re-expand these products in splines, thus casting the original problem into a linear 
system of equations. The accuracy of this procedure was checked by increasing the number 
of splines and a fast convergence was observed.  

For the discrete part of the spectrum in the TDEXX approximation 
we have extracted both excitation energies and oscillator strengths from 
the position and the hight of the peaks of ${\rm Im}\chi$. 
In the RPA and the AEXX the kernels are frequency independent and thus excitation 
energies and oscillator strengths can be directly obtained from the eigenvalues and 
eigenvectors of matrix $\mathbf{V}$, defined in Eq. (\ref{vmatris}).
We will now describe the method that we used in the continuos part of the spectrum.

Once the radius $r_{\rm max}$ is fixed the system studied is actually an 
atom in a box. It turns out, however, that the discrete positive
energy states provide a good description of the true continuum orbitals 
at the same energy but with a different normalization. To get the correct normalization 
the box states only need to be multiplied with a local density of states factor.
By choosing an appropriate mesh both bound and continuum orbitals are well described 
within the same basis.
\begin{figure}
\includegraphics[width=8.5cm, clip=true]{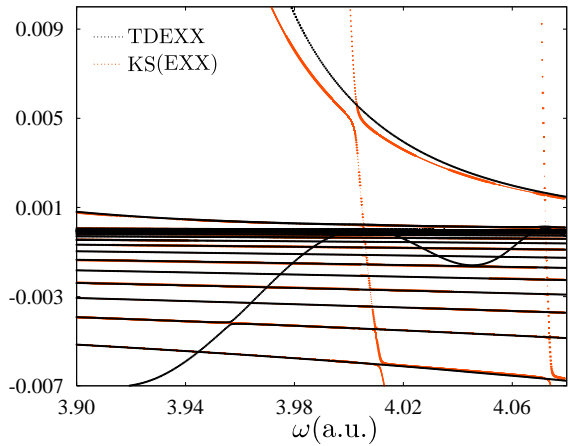}
\caption{The eigenvalues of $\chi_s$ and $\chi$ for Be close the 
first two inner-shell excitations. Two eigenvalues of $\chi_s$ pass 
through zero. As a consequence, $\chi$ in TDEXX has zero eigenvalues 
at the same frequencies but with a different $\w$-behavior.}
\label{bepeak2}
\end{figure}
In our calculation we have used a mesh which starts as cubic and then 
smoothly changes to a linear mesh, according to the formula
\be
\frac{kr^3}{k+r^2}.
\ee

The photoionization cross sections have been obtained by 
first interpolating the discretized ${\rm Im}\chi_{s}$ up to large 
energies, which is possible since it is a smooth function of $\w$ 
for every $\vr$ and $\vr'$. The real part of $\chi_{s}$ was 
then obtained through the Kramers-Kronig relations. 
This was done for every continuum channel. Since the bound states are not
coupled to the continuum in the KS system, the discrete part of 
$\chi_{s}$ can simply be added to form the total non-interacting response 
function of the atom. The resulting response function is then used to solve Eq. 
(\ref{rpafxc}) and to obtain the interacting continuum. 
The accuracy of this method has been checked by calculating 
the RPA spectra of some atoms and comparing them with 
those of other works.\cite{stenerbe,stenerne}
\section{Results and Discussion}
In this section we present numerical results on the EXX kernel for the Be and 
Ne atoms. The photoabsorption spectra calculated using this kernel as well 
as its adiabatic counterpart are presented and compared. We begin with 
a study of the vanishing eigenvalues of the KS linear density 
response function $\chi_s$. This analysis will be useful to 
interpret and understand the structure of the EXX kernel in the
frequency domain.
\subsection{Vanishing eigenvalues of $\chi_s$}
Mearns and Kohn,\cite{mearns} showed that $\chi_s$ 
can have nontrivial vanishing eigenvalues for finite systems at 
certain frequencies.  Physically this means that there is a 
monochromatic perturbation which differs from a constant, and which 
yields a vanishing density response. Note that this is true only if the 
perturbation is not switched on at a particular time. If that is the 
case there is always a finite density response.\cite{robertkey} 

The condition for having a vanishing density response $\delta n$, at 
frequency $\w_0$, can be expressed as 
\be
\delta n(\vr)=\sum_{q}\frac{f_q(\vr)}{\w_0^2-\w_q^2}\int d \vr' 
f^{*}_{q}(\vr')\zeta(\vr')=0,
\label{noll}
\ee
where $\zeta$ is an eigenvector of $\chi_s(\w_0)$ corresponding to an 
eigenvalue equal to zero and describes the spatial part of the monochromatic 
perturbation. With only one occupied orbital like, e.g., in He, the $f_q$-functions are 
linearly independent. In that case $\delta n$ vanishes only if the 
integral in $\vr'$ is zero for every $q$, a condition which implies 
$\zeta(\vr)$ to be independent of position. On the other hand, with more than one 
occupied level the $f_q$-functions are linearly dependent and hence it is, 
in general, possible to find a $\zeta$, different from a constant, that 
fulfills Eq. (\ref{noll}). 

The Be atom has two occupied levels and is therefore the simplest 
closed-shell spin compensated atom for which a nontrivial zero 
eigenvalue can occur. We have diagonalized $\chi_s(\w)$ (in the EXX 
approximation) for the $l=1$ excitation channel of the Be atom in a large finite box 
(40 a.u.) and found a number of eigenvalues which pass 
through zero. The first zero can be found around 2 a.u., an energy 
far above the first ionization threshold. At higher frequencies, a 
set of zeros are found close to the excitation energies of the inner 
$1s$-shell. The eigenvalues of $\chi_s$ for frequencies in the 
range of the $1s\rightarrow 2p$ and 
$1s\rightarrow 3p$ transitions are displayed in Fig. \ref{bepeak2}. 
We note that close to the $1s\rightarrow 2p$ transition (3.948 a.u.)  
there is an eigenvalue which crosses zero at 4.005 a.u.,
and that close to the $1s\rightarrow 3p$ transition (4.058 a.u.) a second 
eigenvalue crosses zero at 4.072 a.u.. We have in fact observed that there 
is one vanishing eigenvalue close to every inner-shell transition. 

The Ne atom has three occupied shells and, therefore, a larger number of vanishing 
eigenvalues are expected. Indeed, vanishing eigenvalues where found
already at low frequencies before the first ionization threshold. 
We did, however, not find any vanishing eigenvalues before the first excitation 
energy, in agreement with the findings in Ref. \onlinecite{mearns}.

For the numerical calculations of the eigenvalues the atom has been confined 
to a finite box, albeit large. The response function is then a sum of discrete 
transitions and it is real almost everywhere. In the limit of an infinite box 
$\chi_{s}$ acquires a finite imaginary part at frequencies in the continuum. 
The zero eigenvalues may then also acquire a finite imaginary part.

As seen from Fig. \ref{bepeak2} the eigenvalue of $\chi_s$ which 
passes through zero is proportional to $\w^{2}-\w_0^{2}$ for 
frequencies close to $\w_{0}$. 
The inverse of $\chi_s$ will, therefore, diverge as a simple pole. 
If the sum $R_{V}+R_{\S}$ in Eq. (\ref{fxekv}) has a 
finite component along the eigenvector with zero eigenvalue 
it may have severe consequences for the calculation of $f_\x$, 
as we will see shortly.

A natural way of handling problems associated with zero eigenvalues of matrices which need to
be inverted is so called singular value decomposition. This means that one finds the eigenvector corresponding to the zero eigenvalue and then invert the matrix in the subspace orthogonal to that eigenvector. Notice, however, that eigenvalues of $\chi_s$ are frequency dependent and 
projecting out states within a limited range of frequencies will result in abrupt changes in the 
spectrum at the borders of the frequency range. And we can find no physical motivation for such 
drastic manual changes.
\begin{figure}
\includegraphics[width=7.5cm, clip=true]{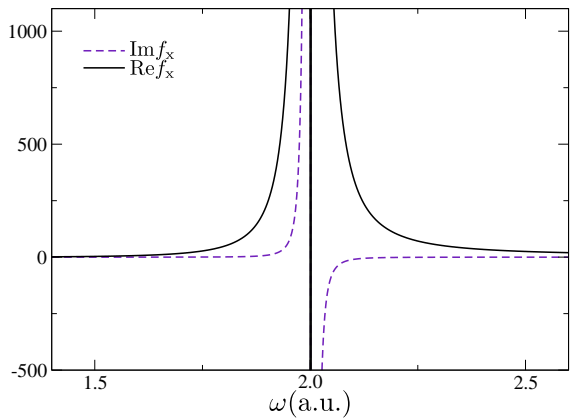}
\caption{The real an imaginary part of $f_\x$ for Be around the first 
singularity.}
\label{bepeak3}
\end{figure}
\subsection{The EXX kernel}
\begin{figure*}
\subfigure[]{\label{bepeaka}
\includegraphics[width=8.5cm, clip=true]{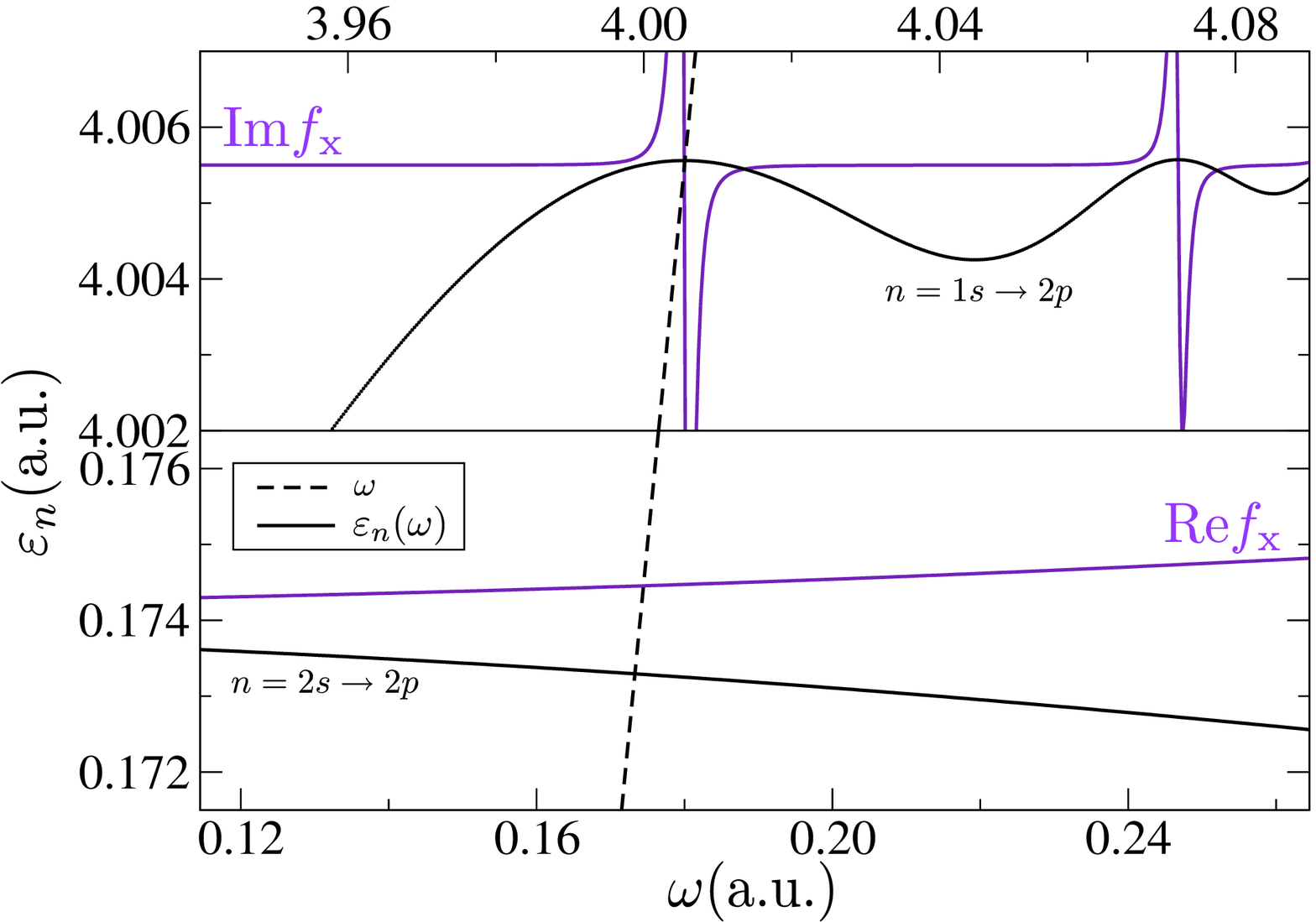}}
\subfigure[]{\label{bepeakb}
\includegraphics[width=9.0cm, clip=true]{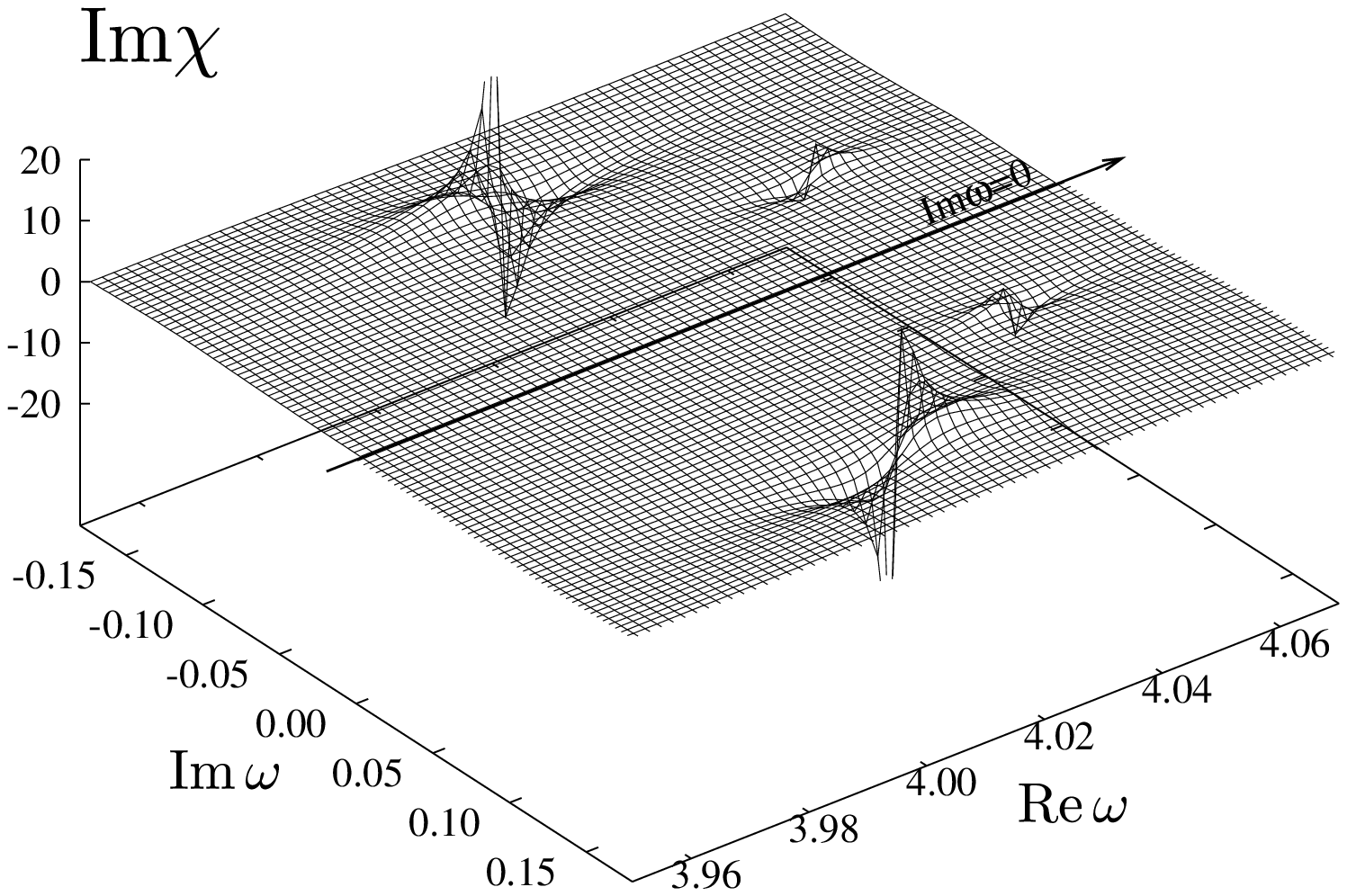}}
\caption{(a) Two different eigenvalues of the matrix $\mathbf{V}$ for 
Be. The upper/lower panel shows an eigenvalue corresponding to the 
first inner/outer-shell excitation. The intersection with the dashed 
line gives the positions of the new resonances. For the inner-shell 
eigenvalue this occurs exactly where $f_\x$ has a pole. At that value the 
oscillator strength is, however, zero. Two more solutions are found but for 
complex $\w$. (b) The imaginary part of $\chi$ is plotted on the 
complex plane. Close to every inner-shell transition two poles are found in 
the complex plane.}
\label{bepeak}
\end{figure*}
As noted previously,\cite{argunn,hvb2} the XC kernel is not uniquely 
defined by Eq. (\ref{fxceq}). Given 
an $f_\xc$ one can always add two arbitrary functions $g_1(\vr,\w)$ 
and $g_2(\vr',\w)$
without changing the results for $\chi$. This originates in the fact 
that the potential is determined 
up to the addition of a purely time-dependent function and that the 
density variations must integrate to zero. When we discuss 
$f_\x$ and
its dependence on the frequency we instead consider the quantities  
$f^{qq'}_\x(\w)=\bra q|f_\x(\w)|q'\ket$, which are unique since the 
excitation functions integrate to zero and thus remove the effect of 
adding functions of the form $g_1$ and $g_2$. 

Let us first consider the case of the He atom, or for that matter, 
any other spin-compensated 
two-electron system. In this case $f_\x$ equals minus half the 
Coulomb potential $(-\frac{1}{2} v)$. This can be seen either by 
noting that the EXX potential becomes $-\frac{1}{2}\int nv$ and hence 
the functional derivative can be taken explicitly, or after some 
manipulations of the diagrammatic expression in Eq. (\ref{fxekv}). 
In the latter case it is important to impose the LSS equation 
as well as the additional constraint\cite{kli} 
\bea
\bra\varphi_N|v_\x-\S_s^\x|\varphi_N\ket=0 \quad \mbox{if and only 
if} \quad\lim_{r\to\infty}v_{\x}(r)=0,&&\nn
\eea
where $N$ denotes the highest occupied orbital. We thus see that for 
He the EXX kernel is energy independent. 

Let us now consider systems with more than one occupied 
orbital. The Be atom has two filled closed shells - the inner
1s-shell and the outer 2s-shell - and a full numerical solution
is therefore required. The kernel for the $l=1$ channel was calculated 
with a finite box radius of 40 a.u.. At low frequencies $f_\x$ has a weak 
frequency dependence and is purely real. At larger frequencies, however, 
$f_{\x}$ develops a strong frequency dependence and around 2 a.u. the 
kernel diverges. In Fig. \ref{bepeak3} we show the real and imaginary 
parts of $f_\x$ around this point. The divergence is typical of a double pole 
with the real part given by the difference between a term $1/(\w-\w_0)^2$ 
and a squared delta function, $\delta(\w-\w_{0})^{2}$.  A small asymmetry 
around the pole can be observed which indicates the presence of an
additional simple pole. The exact location of the pole can be traced 
back to the first vanishing eigenvalue of $\chi_s$ (see discussion 
in the previous section). The pole structure of $f_\x$ shows that 
$R=R_V+R_\S$ does not compensate for the zero of $\chi_s$.
\begin{figure*}
\subfigure[]{\label{nepeaka}
\includegraphics[width=8.5cm, clip=true]{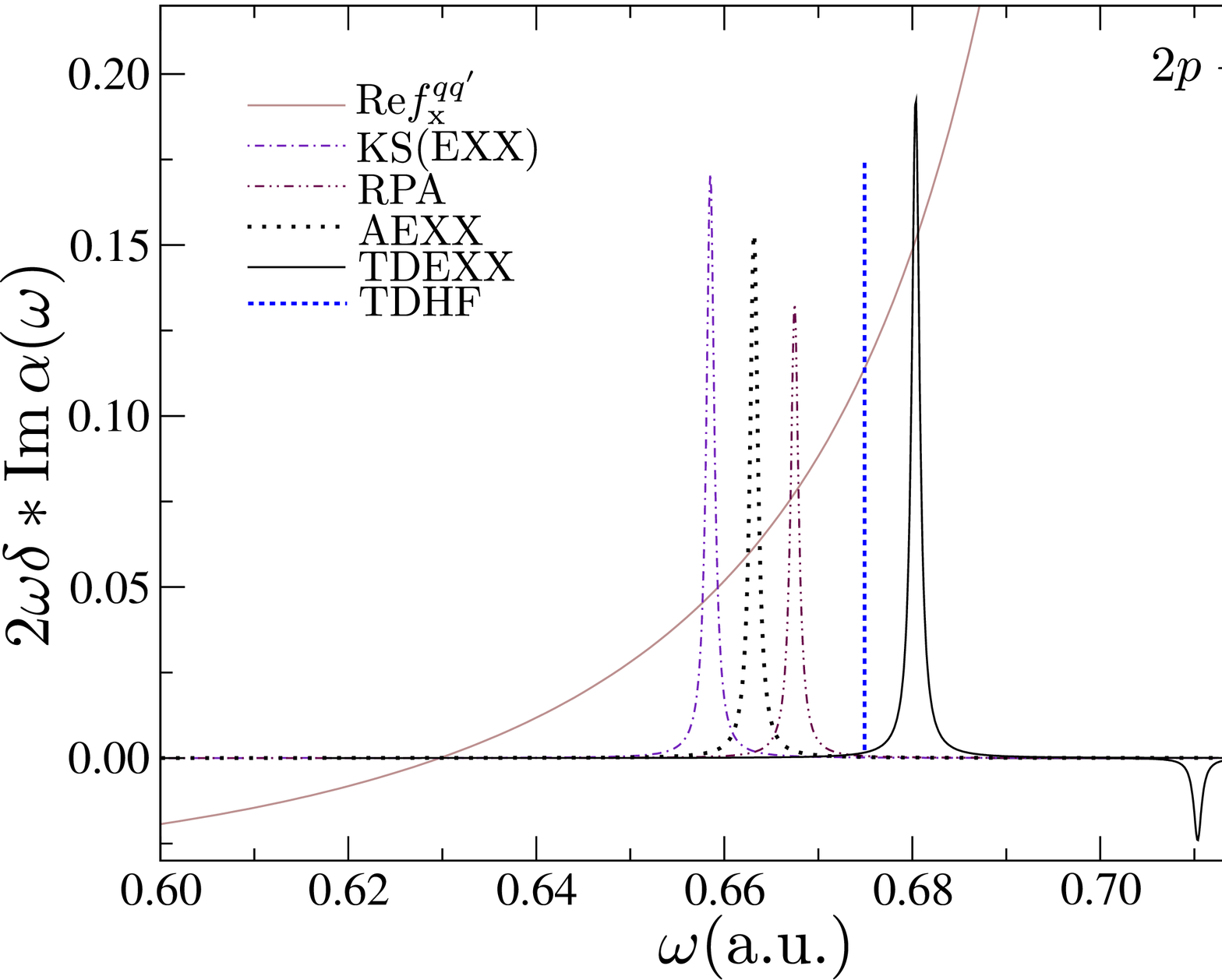}}
\subfigure[]{\label{nepeakb}
\includegraphics[width=8.5cm, clip=true]{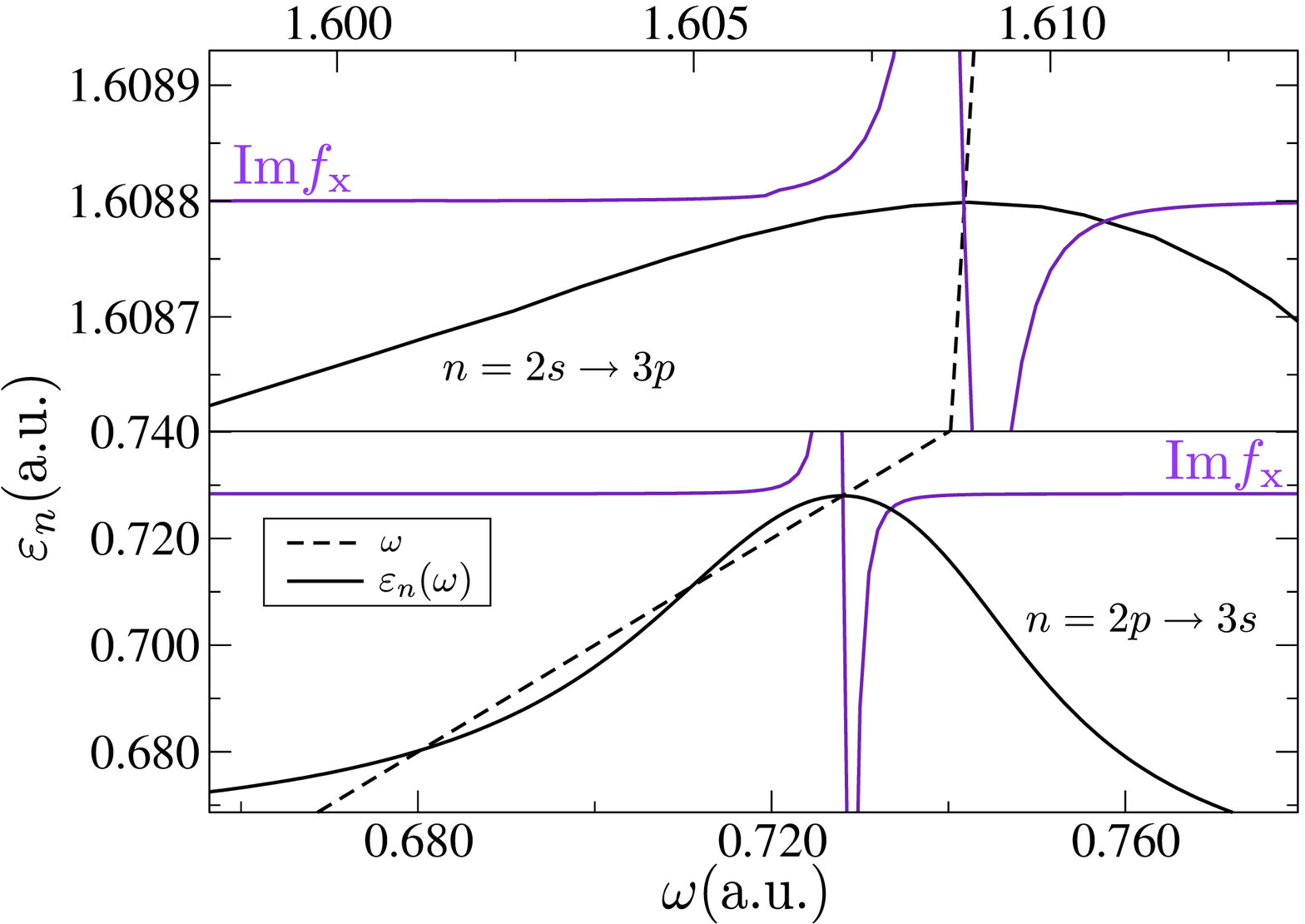}}
\caption{(a) The absorption spectrum of Ne around the $2p\rightarrow 
3s$ transition. The adiabatic approximation to $f_\x$ moves the RPA 
value in the wrong direction since the real part of $f_\x$ is 
negative at $w=0$. The strong frequency dependence of $f_\x$ moves the 
peak closer to the TDHF values as it should. Also a small negative peak is 
generated close to the pole of $f_{\x}$ at 0.73 a.u.. (b) The lower panel shows the eigenvalue corresponding to the 
$2p\rightarrow 3s$ transition. Three intersections are found. Two of 
these are seen in the left figure while the third has a zero oscillator strength. }
\label{surf}
\end{figure*}

Continuing to higher frequencies more singularities in 
the kernel are observed. Near the $1s\rightarrow 2p$ transition in $\chi_s$ 
a second pole in $f_{\x}$ develops. As a matter of fact we find a pole in 
$f_\x$ close to every $1s\rightarrow np$ transition in $\chi_s$ (see the upper 
panel of Fig. \ref{bepeaka}). These poles have the same structure as 
the one at 2 a.u. and are also confirmed to originate from the vanishing 
eigenvalues of $\chi_s$. The vanishing eigenvalues corresponding to the first 
two inner-shell excitations are displayed in Fig. \ref{bepeak2}. 
The double and simple pole structure in $f_{\x}$ is expected since, in order 
to calculate the kernel, $\chi_s$ has to be inverted twice, see Eq. 
(\ref{fxceq}). Let us call $\zeta_{l}(\vr,\w)$ the eigenvector of 
$\chi_{s}(\vr,\vr',\w)$ with eigenvalue $\l_{l}(\w)$. Then from the 
equation $\chi_{s}f_{\x}\chi_{s}=R$ the kernel can be expressed as
\be
f_{\x}(\vr,\vr',\w)=\sum_{ll'}\frac{\bra\zeta_{l}|R|\zeta_{l'}\ket}
{\l_{l}(\w)\l_{l'}(\w)}\zeta_{l}(\vr,\w)\zeta^{*}_{l'}(\vr',\w).
\label{fx}
\ee
If we denote by $\w_{k}$ the frequency for which there is an
eigenvalue $\l_{k}$ equal to zero, we have that 
$\l_{k}\sim\w^{2}-\w_{k}^{2}$ for $\w\sim\w_{k}$ and Eq. (\ref{fx}) 
can be cast in a more transparent form
\be
f_{\x}(\w)=f^0(\w)+\sum_k \left\{\frac{f^{(1)}_k(\w)}{\w^2-\w^2_k}  
+\frac{f^{(2)}_k(\w)}{(\w^2-\w^2_k)^2}\right\} 
\ee
where the quantities $f^0$, $f^{(1)}_k$ and $f^{(2)}_k$ are weakly dependent on 
$\w$, and the $\vr$ and $\vr'$ dependence has been suppressed. 
Notice that the object $\chi_sf_\x$, the basic quantity needed to 
calculate the full response function $\chi$, only has simple poles at the same 
frequencies. 

The kernel for the Ne atom has an even larger number of poles due 
to the larger number of inner-shells and thus a larger number of vanishing 
eigenvalues of its $\chi_s$. All poles are of the same structure as those discussed 
for Be, but the first pole appears already between the first and second resonance in 
$\chi_s$, i.e., below the ionization threshold of Ne. 

As can be seen from Eq. (\ref{fx}) the $f_{q}$-representation is not 
well suited for studying the singular behavior of the kernel in real space. 
The elements of the matrix $\mathbf{V}$ in Eq. (\ref{vmatris}) will all 
have the pole structure described above. When the
matrix is diagonalized, however, this pole structure survives in only 
one of the eigenvalues, whereas the remaining ones appear to have a 
smooth frequency dependence but still affected by the pole giving rise to an 
oscillatory behavior. In Figs. \ref{bepeaka} and \ref{nepeakb} two different smooth 
eigenvalues of this matrix are shown for Be and Ne respectively. The eigenvalue 
corresponding to the $2s\rightarrow 2p$ transition in Be is weakly dependent on 
the frequency and hence the intersection with the line $\w$ giving the 
location of the new excitation energy is only slightly shifted from that of the adiabatic 
approximation. The new position in the TDEXX approximation is in 
better agreement with the corresponding excitation energy of the TDHF 
approximation as compared to the same position in the adiabatic case, 
see Tab. \ref{beneexc}. The position of the first discrete excitation in 
Ne is, however, strongly modified by the frequency dependence as is 
shown in Fig. \ref{nepeaka}. The kernel of the adiabatic 
approximation moves the peak to a lower frequency as compared to the RPA 
while the TDHF result lies at a higher frequency (0.674 a.u.). The frequency 
dependence of the kernel leads to the opposite effect and moves the peak 
even a little beyond the true TDHF value. At a somewhat larger 
frequency the EXX kernel has a pole and the eigenvalue corresponding 
to the $2p\rightarrow 3s$ excitation is seen to oscillate (see lower panel 
of Fig. \ref{nepeakb}). The maximum of the oscillation occurs exactly at the 
location of the pole. Apart from the desired solution, i.e., the 
crossing with the $\w$-line at 0.680 a.u., there are two more 
crossings. The first of these corresponds to the small negative peak 
found in the spectrum (at 0.71 a.u.) and the next has zero oscillator strength. 
Clearly, the negative peak is unphysical but we expect that the ability of 
the TDEXX to describe the spectrum rapidly decreases as we approach 
the pole. 

Since there is a pole close to every inner-shell excitation 
the eigenvalues oscillate in these regions. The eigenvalue 
corresponding to an inner-shell excitation gives only one solution 
exactly at the pole as seen in the upper panels of Figs. 
\ref{bepeaka} and \ref{nepeakb}. This is true for all inner-shell 
transitions in Be and Ne and seems to be a general behavior. The 
oscillator strength is zero at these points leading to a complete 
disappearance of the peaks from the spectrum as also seen, e.g., at 
third crossing corresponding to the first excitation in Ne as discussed above. 
Because the sum rule is obeyed the missing oscillator strength is expected to 
be transferred to excitations of the outer shell. This conjecture is based on having the correct 
analytic structure of the response function with a Lehmann 
representation leading to the fact that the large frequency behavior 
of the response is determined by the sum of the different oscillator 
strengths. Unfortunately, in many cases the three solutions at the 
real frequency axis close to the double pole structure, which we 
discussed above, instead leads to one solution at the real axis and 
two more complex solutions one of which is in the upper half plane and 
the other in the lower. 
\begin{table*}[t]
\begin{ruledtabular}
\begin{tabular}{llllllll}
Trans.&KS(EXX)&RPA&AEXX&TDEXX&TDHF\footnotemark[1]&KS(Exact)\footnotemark[2]&Exp.\footnotemark[4] 
\\
\hline
Be&&&&\\
1s$\rightarrow$2p& 3.948& 3.959& 3.956 
&4.005\footnotemark[3]&4.346&4.017 & 4.243 \\
1s$\rightarrow$3p& 4.058& 4.060 & 4.059 
&4.072\footnotemark[3]&4.646   &4.153 & 4.461  \\
2s$\rightarrow$2p& 0.131&0.203 &0.177&  0.176    &0.176   & 0.133& 
0.194 \\
2s$\rightarrow$3p& 0.241  &0.255 &0.247& 0.247    &0.247 &0.269& 
0.274 \\
\\
Ne&&&&\\
2s$\rightarrow$3p& 1.604  &    1.608       & 1.607&    
1.609\footnotemark[3]  &        -      &1.542    &1.674 \\
2s$\rightarrow$4p&1.667   &     1.668      & 1.667&   1.669\footnotemark[3]    &       
-        &1.602  &  1.731\\
2p$\rightarrow$3s& 0.659 &0.667 & 0.663&0.680 &  0.674  &0.612  
&0.619 \\
2p$\rightarrow$4s& 0.779 &0.781  & 0.780&0.783  &  0.782  &  0.725 &  
0.727
\label{beneexc}
\end{tabular}
\end{ruledtabular}
\footnotetext[1]{From Ref. \onlinecite{tdhfbe} and \onlinecite{tdhfne}}
\footnotetext[2]{From Ref. \onlinecite{umr}}
\footnotetext[3]{Transition calculated assuming no inter-shell 
coupling.}
\footnotetext[4]{From Ref. \onlinecite{nist}}
\caption{The two first discrete excitation energies from the $2s$ and 
$1s$ shell of Be and the $2p$ and $2s$ shell of Ne. Different 
approximations for the kernel is used in conjunction with the EXX 
potential for the ground state. A column with the eigenvalue 
differences calculated using the exact KS potential of Umrigar et. 
al.\cite{umr} is also presented. All values are in a.u..}
\end{table*}
Thus the proper analytic structure of the response is destroyed and 
the fact that the TDEXX response has the correct large frequency 
behavior does not guarantee that the full spectrum contains the 
correct sum of oscillator strengths. Part of the spectrum is actually 
missing which is a clear failure of the theory. In Fig. 
\ref{bepeakb} $\chi$ clearly exhibits two peaks located symmetrically 
on opposite sides of the real axis. Apart from the solution with zero 
oscillator strength there are thus two more solutions at complex 
frequencies ($\w$). 

Finally, before further discussing the spectrum we will make a 
small digression and investigate the eigenvalues of $\chi$. In Fig. 
\ref{bepeak2} the eigenvalues of both $\chi_s$ and $\chi$ are 
presented in a frequency range where there are two vanishing 
eigenvalues of $\chi_s$. We see, however, that also $\chi$ have 
vanishing eigenvalues at the same points. We also see that the 
eigenvalues of $\chi$ only touches the zero intensity axis whereas 
the eigenvalues of $\chi_s$ cross this axis.
This can be explained by studying
\be
\chi=\frac{\chi_s}{1-v\chi_s-\chi_s^{-1}R}
\ee
obtained by using the equation $f_\x=\chi_s^{-1}R\chi_s^{-1}$. Our calculations 
show that the matrix $R$ does not have a zero eigenvalue where 
$\chi_s$ has. Consequently, by writing this equation in the basis of 
the eigenvectors of $\chi_s$, we 
immediately see that $\chi$ behaves as $(\w^{2}-\w_0^{2})^2$ close to $\w_0$. 
There is no reason to believe that the Coulomb interaction would 
change the simple zeros of $\chi_s$ to double zeros. We therefore
expect also $\chi$ to have simple zeros although shifted relative to 
those of $\chi_s$.
From the definition of the exact XC kernel,
\be
f_\xc=\chi^{-1}-\chi_s^{-1}-v,
\ee
$f_\xc$ should have simple poles where $\chi_s$ and $\chi$ have 
vanishing eigenvalues. The double pole structure
found within the EXX approximation is thus an artifact of this 
theory. 
\subsection{Discrete excitation energies}
Before the first ionization threshold the atomic photoabsorption 
spectra consists of a set of 
discrete transitions. In Tab. \ref{beneexc} the first two discrete 
transitions are presented for 
Be and Ne. A comparison is made between the AEXX, TDEXX, TDHF and the 
RPA 
as well as the KS eigenvalue differences of the exact\cite{umr} and 
the 
EXX potentials. The latter potential was also used when calculating 
$\chi$ in the aforementioned
approximations. 

The TDEXX approximation is expected to give results close to TDHF. In 
this sense the outer-shell transitions of Be are well described already 
in the AEXX approximation. By also accounting for the frequency dependence 
the results improve even further. For the outer-shell excitations of Ne 
dynamic effects in $f_\x$ are crucial. The AEXX approximation reduces 
the values of the RPA whereas the TDHF results are larger than those 
of the RPA. The frequency dependence corrects this tendency yielding results
in good agreement  with TDHF, although somewhat over-estimated. The 
effect of the kernel on the first excitation energy of Ne is illustrated 
in Fig. \ref{nepeaka}. The results of TDHF and hence of TDEXX differ 
markedly from the experimental values. By looking at the exact KS 
eigenvalue differences it can be concluded that the error is mainly related 
to the effect of the ground state potential. The kernel has only a small 
effect on the outer shell excitation energies as was previously 
pointed out by Petersilka et. al..\cite{pgb} 
\begin{figure*}[t]
\subfigure[]{\label{bespe}\includegraphics[width=8.5cm, 
clip=true]{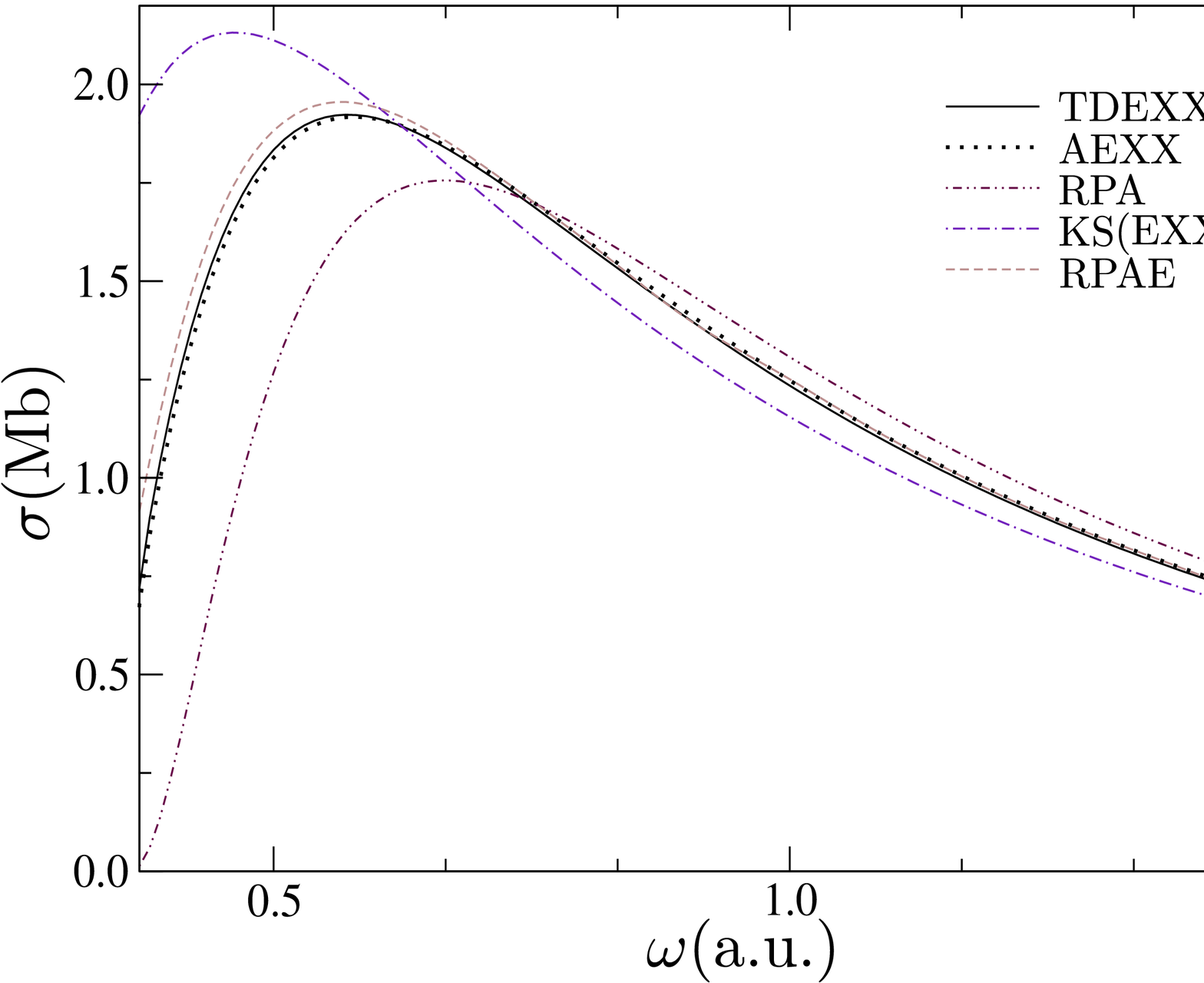}}
\subfigure[]{\label{befano}\includegraphics[width=8.5cm, 
clip=true]{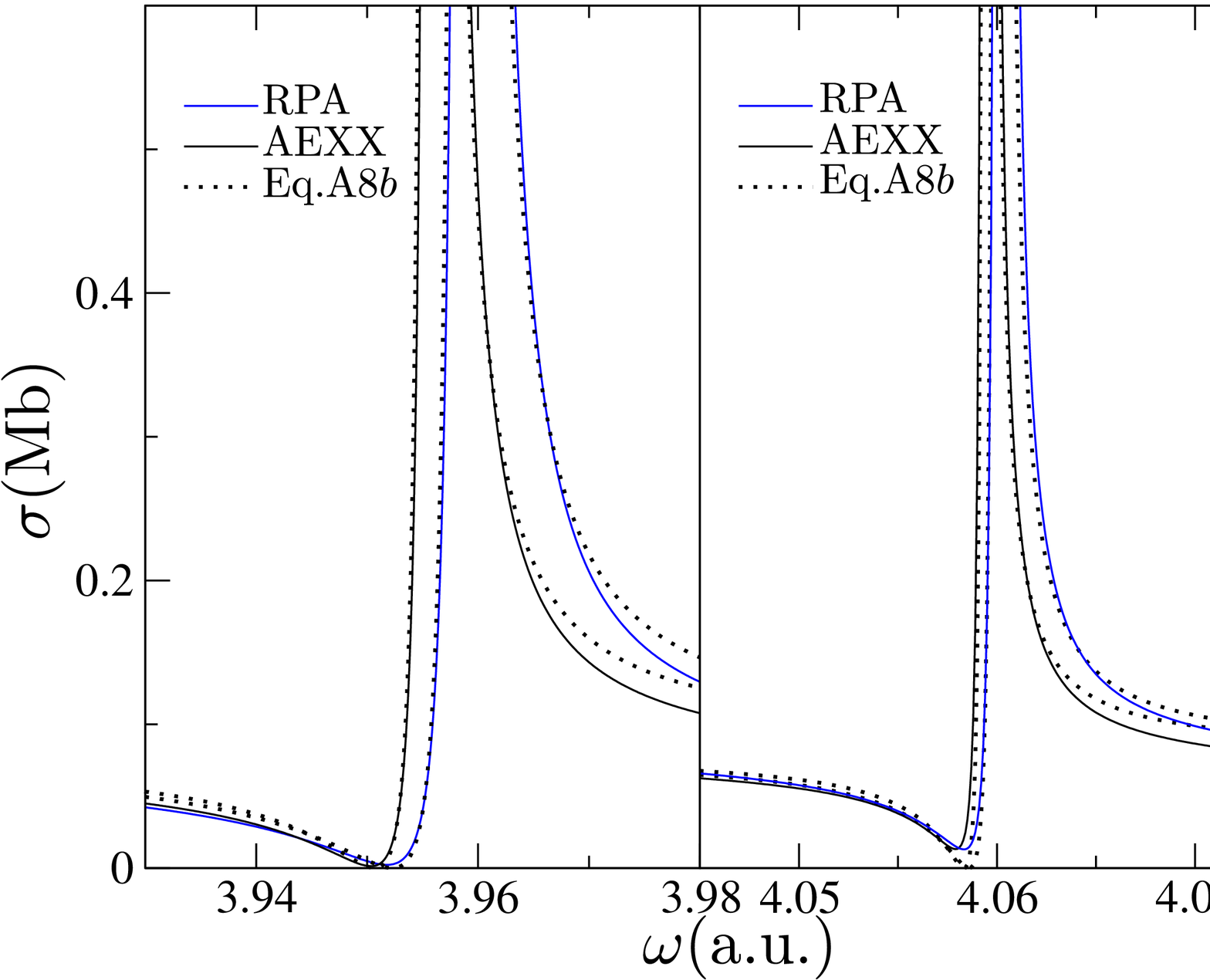}}
\caption{(a) Photoionization cross section for Be after the first 
ionization threshold. 
(b) First two Fano resonances resulting from the $1s\rightarrow 2p$ 
and $1s\rightarrow 3p$ transitions.}
\label{bespec}
\end{figure*}

A description of inner-shell excitation energies in terms of density 
functional eigenvalues is bound to fail.
In Tab. \ref{beneexc} we see that the true core excitation energies 
are largely under-estimated by 
even the exact KS eigenvalue differences as compared to the true 
excitation energies. An adiabatic 
kernel is unable to correct these errors as confirmed by 
the examples (RPA and AEXX) shown in the table. A strong frequency 
dependence in the kernel might, however, improve the situation considerably.
The vanishing eigenvalues of $\chi_s$ result in a strong frequency 
dependence in the EXX kernel near these 
excitations. As we saw in Sec. IIIC this frequency dependence is, 
however, too strong and even produces the 
wrong analytic structure of the response.
The matrix $\mathbf{V}$ contains separate diagonal blocks for 
different excitation 
channels and off diagonal blocks for the coupling between them.  
Notice, however, that this
separation is not entirely well defined since part of the 
inter-channel coupling is already contained in 
$f_\xc$. By including $f_\x$ and diagonalizing only the block 
containing the inner-shell excitations we 
obtain some improvement for the position of these.  Accordingly, 
for Be we have diagonalized the $1s\rightarrow np$ 
block and for Ne we have diagonalized the $2s\rightarrow np$ block. 
The results are marked $c$ in
Tab. \ref{beneexc}. The most important observation here is the fact 
that by means of this somewhat ad-hoc procedure we can obtain results 
for those inner-shell excitations which completely vanish in the full 
treatment. 

The calculation of the inner-shell transitions was done with a finite 
box size and the excitation energies
where extracted from the location of the delta peaks in ${\rm 
Im}\chi$. In reality these peaks will be 
somewhat shifted and broadened forming a so-called Fano resonance due to 
the coupling to continuum channels. In the next section
we will present results for these resonance structures.
\subsection{Photoionization cross sections and \\ Fano resonances}
\label{consp}
The $2s\rightarrow 2p$ photoionization cross section of Be is 
presented in Fig. \ref{bespe} and different approximations are 
compared. The experimental result has not been 
included since in this region the spectrum is dominated by the $2p,ns$ 
and $2p,nd$ double excitation resonances, which cannot be captured by 
the approximations studied here. 
The figure shows that the AEXX and the TDEXX approximations are 
almost indistinguishable and also very close to the RPAE
results of Amusia et. al.\cite{ambe} We remind the reader that the 
RPAE approximation is identical to linearized TDHF. The frequency 
dependence of the kernel is weak up to around 2 a.u. where the first  
singularity of $f_\x$ occurs. This strong frequency dependence of 
$f_\x$ has, however, only a small effect on the spectrum in the 
studied region. 

\begin{figure*}[t]
\subfigure[]{\label{nespe}\includegraphics[width=8.5cm, 
clip=true]{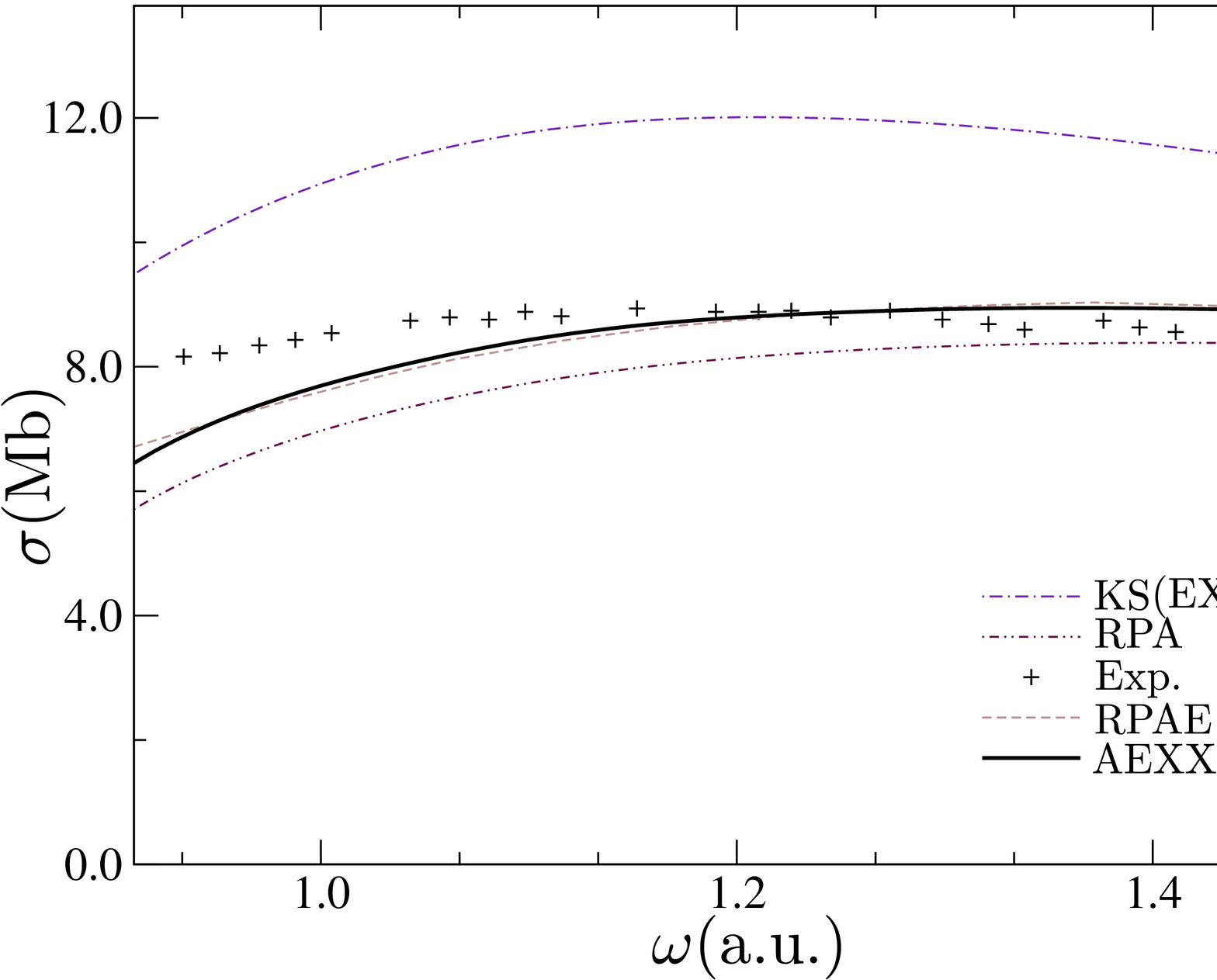}}
\subfigure[]{\label{nefano}\includegraphics[width=8.cm, 
clip=true]{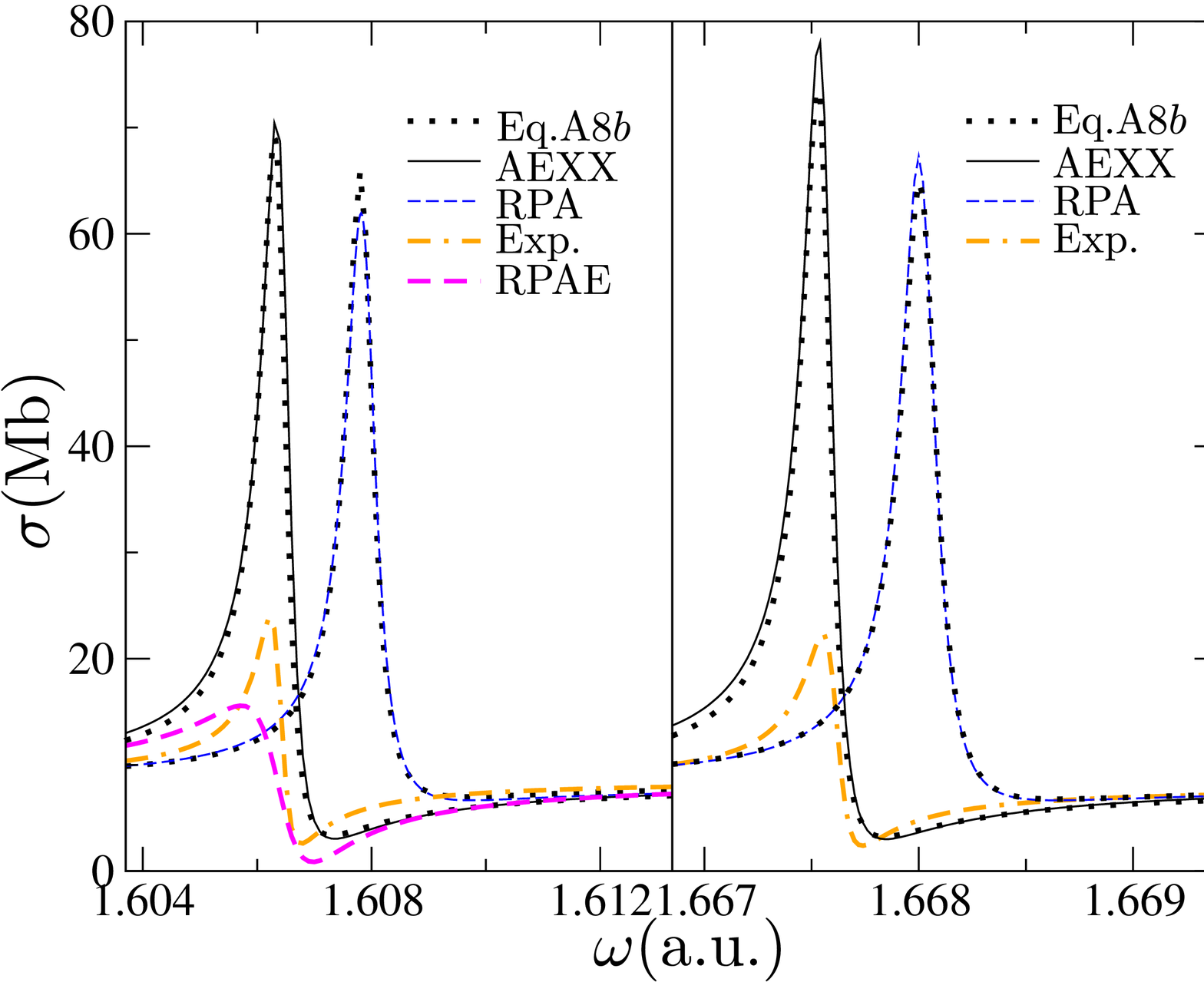}}
\caption{(a) Photoionization cross section for Ne just after the 
first threshold. 
(b) First two Fano resonances arising from the $2s\rightarrow 3p$ and 
$2s\rightarrow 4p$ transitions.
Notice that the experimental and the RPAE results have been shifted to 
allow for a better comparison of widths and $q$-parameters.}
\label{nespec}
\end{figure*}

The photoionization cross section after the first ionization 
threshold of Ne is presented in Fig. \ref{nespe}. 
It contains two ionization channels, $2p\rightarrow$ to continuum $s$ 
or $d$. The AEXX agrees very well with 
the RPAE results.\cite{amne} There is also a fairly good agreement 
with experiment.\cite{chan} In the figure we have deliberately not
presented results obtained by including the full energy dependence of 
$f_\x$. In the energy region covered
by Fig. \ref{nespe} the non-interacting response function of Ne has a 
number of zero eigenvalues producing violent 
structures in $f_\x$. This unphysical behavior results in a series 
of strong peaks also in the full response. 
Thus including the energy dependence of $f_\x$ results in absorption 
spectra with no resemblance to the 
spectra in Fig. \ref{nespe}. These strong structures are of course 
obtained using discrete splines which give only a discrete sampling 
of the continuum. It could be possible that calculating $f_\x$ using true continuum 
functions would somehow alleviate this problem even though it will not go 
away completely. The zero eigenvalue below 
the ionization threshold does produce unphysical structure 
in the absorption spectra of Ne. In this case one cannot blame the 
unphysical spectrum on the lack of a proper continuum, and we are 
convinced of the futility in going through the trouble of calculating 
proper continuum functions. 

At higher frequencies, Fano resonances due to excitations of inner-shell 
electrons occur. In the Appendix we present a derivation of the Fano 
profile formula for excitations with single-particle character 
starting from the linear density response function within adiabatic TDDFT.
The Fano parameters $q$, $\G$, and $\rho^2$ are there given expressions in 
terms of the adiabatic kernel $f_\xc$, see Eq. (\ref{parameters}) and Eq. 
(\ref{rho}). In  Figs. \ref{befano} 
and \ref{nefano} we compare spectra calculated either 
directly from Eq. (\ref{rpafxc}) using an interpolated $\chi_{s}$ or 
by employing the analytic method described in the Appendix. Within the 
analytic approach we have for both Be and Ne ignored the coupling between 
different discrete states via the continuum (which corresponds to 
ignoring the off-diagonal elements of the matrix $\mathbf{M}$ in Eq. (\ref{offdiag})).
For Ne we have also assumed that we can ignore the coupling between the $s$ 
and $d$ continua, which can be reached from the $2p$-shell. The Figs. \ref{befano} 
and \ref{nefano} contain results of both the RPA and the AEXX approximation. As seen, 
the two different ways of calculating the spectra around a resonance give almost 
identical results. This verifies the validity of the assumptions made above.

The two first inner-shell resonances of Be are very sharp with a zero 
on one side of the resonance due to the coupling to only one continuum.  
The very large $q$-parameter is accompanied with a very small width, 
$q\!\approx\!220$ and $\G\!\approx\!1.6$ meV in the RPA and 
$q\!\approx\!320$ and $\G\!\approx\!0.86$ meV in the AEXX 
approximation, for the case of the $1s\rightarrow 2p$ resonance. In Ref. 
\onlinecite{stenalda} the same parameters were calculated using the 
ALDA kernel giving similar results for the $q$-parameter. The 
width ($\G$) from the AEXX approximation is, however, a factor of two smaller 
than those from both the RPA and the ALDA. Experimental results have 
in this case not been found. 

In Ne the discrete excitations of the $2s$-shell couples to both the 
$p$ and the $d$ continua.
The minima on one side of the resonances therefore never reaches 
zero, which means that $\rho^2<1$. This parameter is for the first resonance 
$\rho^{2}\approx 0.17$ in the RPA and $\rho^{2}\approx\!0.60$ in the 
AEXX approximation. Since the experimental value is $0.70$ the AEXX 
approximation improves over the RPA. 
The widths are $\G\approx\!15.4$ meV in the RPA and 
$\G\approx\!14.4$ meV in the AEXX approximation. Again, the value of the 
AEXX approximation is in quite good agreement with the experimental result of 
13 meV. The $q$ parameter differs, however, substantially from the 
experimental values. The RPA gives 6.3 and 
the AEXX approximation 3.5 whereas the experimental value is 1.6. 
In Fig. \ref{nefano} the two first resonances of Ne are displayed. 
We clearly see that the shape is very sensitive to the 
$q$-parameter. Notice that the experimental\cite{codling} and the RPAE results of 
Amusia\cite{amusiacase} are shifted in order to facilitate the comparisons. 
The RPAE approximation yields $q=0.95$ and $\G=34$ meV. The latter is about 
three times larger than the experimental value. 
In conclusion we can say that for the $2p\rightarrow 3s$ resonance of 
Ne neither of the static approximations here give a reasonable account of the 
experimental findings. The deviations to experiment is in many cases much 
larger than the results from different theories. A more sophisticated 
theory is therefore required. Theories which do provide a 
reasonable description of the spectra are wave function based, include many terms 
in a configuration interaction expansion, and can hardly be used on larger systems.
\section{Summary and conclusions}
In the present work we have used the so called exact-exchange
approximation within TDDFT to calculate particle conserving
discrete excitation energies and photoabsorption spectra of a few
spherical atoms. As in several earlier papers on this topic we have
also here relied on a numerical method based on cubic splines as basis
functions for one-electron wave functions, polarizabilities, and the
XC kernel. Clearly, such a numerical method amounts
to a discretization of the continuum in which we have no control over 
the resulting positions of one-electron eigenvalues or electron-hole
excitation energies which are ingredients in the necessary correlation
functions. In order to circumvent such problems we have here
constructed interpolation schemes which work very well for the
description of, e.g., the non-interacting KS response function
and other functions in the continuous part of the spectrum. In 
addition, we present in the Appendix a new way in which the parameters
of the Fano profiles associated with the autoionizing resonances
corresponding to inner-shell excitations can be calculated directly
within adiabatic TDDFT without resorting to a full calculation of the 
entire spectrum. The resulting parameterized spectra are actually very accurate
approximations to the full spectra which was demonstrated
for the RPA and for the spatially non-local EXX kernel evaluated at zero 
frequency (AEXX). But, unfortunately, it has not been true when we have tried to incorporate
the full frequency dependence of the EXX kernel. For reasons
which have been discussed in the paper and will be mentioned again
later on, the violence of that frequency dependence has defied any
reasonable interpolation scheme. As a result we present few
results based on the fully frequency dependent EXX kernel. 

One such result is the calculation of discrete transition energies.
There are really no strong trends in the results.
We here compare our results to the corresponding results from a full time-dependent
Hartree-Fock (HF) calculation. As discussed in the paper, this is
motivated by our belief that the latter calculation is the 'target'
for the TDEXX approximation. It is not motivated by the HF results being 
very close to experiment which they are not.  We find that the rather poor eigenvalue 
differences in the EXX approximation are improved by the inclusion of the Hartree part 
of the kernel (the RPA results). Then, for Ne, the situation deteriorates when the static 
(adiabatic) exchange effects are included just to again improve when the exchange 
effect are treated in a frequency dependent fashion.  The full frequency-dependent 
kernel has, however, a tendency to overcorrect the errors from the RPA. In the case of 
Be the results of the adiabatic approximation instead represent a marked improvement  
on the RPA results and the frequency dependence gives a slight additional improvement. 
In all fairness we must add that these feeble trends are completely overshadowed by 
starting from a better XC
potential like the exact DF potential available for the light atoms or
from the very similar XC potential from a full GW calculation. The
outer transition energies then move much closer to experimental 
results.

In the continuous part of the spectra we have used our new way of 
obtaining the parameters of the Fano line shapes and compared our
results to those of the RPA, to previous calculations using different 
versions of the ALDA, and to experiment. We have found that all
adiabatic kernels including the AEXX give a clear improvement over
the RPA results for those Fano parameters which, roughly, correspond 
to the weights and the widths of the resonances with a slight edge for
the AEXX approximation.  
The fact that the parameter mainly responsible for the height
of the resonances is poorly described by all static approximations 
suggests that the former results are fortuitous. This impression is
further reenforced by observing that a supposedly superior treatment 
including full energy dependence like the RPAE yields results for the 
parameters which are further away from experiment. Turning then to
the density-functional version of RPAE, i.e., the TDEXX the description 
completely breaks down and the resonances disappear as will now be
discussed.         

Perhaps the most important finding of the present work is the violent
frequency variation of the EXX kernel discovered in the vicinity
of energies associated with inner-shell excitations. This, in turn,
is a result of vanishing eigenvalues of the non-interacting Kohn-Sham
response function close to these energies, and the necessity to invert
that response function in order to obtain the resulting EXX kernel -
at least in the way we presently formulate the theory. Although we can
observe a tendency of the EXX kernel to use a strong frequency
variation as a way to mimic the effect of the non-local but frequency
independent three-point vertex of time-dependent Hartree-Fock theory
this frequency behavior is quite unphysical and actually leads to the
complete disappearance from the spectrum of all inner-shell
excitations. We here trace this breakdown of the TDEXX approximation to the
appearance of poles at a finite distance in frequency into the upper 
complex frequency plane thus producing a density response function 
with an incorrect analytic structure. One might think that this
could be an order of limits issue and that the problem might go away 
if a proper continuum was included in the calculations. We doubt the
correctness of this conjecture because the problem is already visible
in the truly discrete part of, e.g., the Ne spectrum. 
In the present approach we implicitly calculate the linear response
to a perturbation which has a sinusoidal variation at all times. It
has been shown by van Leeuwen\cite{robertkey} that the non-interacting response
function is always invertible for perturbations which completely
vanish before a given time. This suggests that a proper TDEXX spectrum
could be obtained by applying a sinusoidal perturbation at one
particular time, propagate the time-dependent EXX orbitals, wait a
time T until all transients have died out and then Fourier
transform the charge density starting from time T to a very large
time. This would amount to some limiting procedure which is quite feasible 
to carry out. It is, however, hard to see that such an approach would lead 
to something different to what we already have obtained.

It should be noted that the inclusion of higher-order correlation
effects like, e.g., screening or relaxation effects will not be a
remedy to the problem. Any procedure for obtaining better XC kernels
based on the variational formulation of many-body theory and the Klein
functional will lead to the linearized Sham-Schl{\"u}ter equation at the
first variation, albeit with a very sophisticated self-energy
containing many physically important correlation effects. One further
variation of the LSS equation in order to obtain the kernel will again reveal
the necessity to invert the non-interacting Kohn-Sham response and the
problem is still there. Only by pure coincidence will the right hand
side of the equation determining the kernel have a zero eigenvalue at
the same frequency as the non-interacting response.
A possible resolution to the problem could be starting from a more
sophisticated and variationally stable functional than the Klein
functional, e.g., the functional of Luttinger and Ward (LW), see Ref. \onlinecite{fot1}. The
additional computational work will be substantial already at the
exchange-only level but it is definitely worth a serious research effort. 
But then, of course, we are no longer working within the TDEXX. The underlying 
$\Phi$-functional will still be at the level of the HF approximation
but it could be hoped that the more complicated LW functional 
would render a kernel within TDDFT which would result in a response 
function much closer to that of TDHF. The latter is expected 
to have the correct analytic structure but it does not produce 
overly impressive results.
\begin{acknowledgments}
The authors would like to thank R. van Leeuwen 
and C.-O. Almbladh for useful discussions.
This work was supported by the European Community Sixth Framework 
Network of Excellence NANOQUANTA (NMP4-CT-2004-500198) and the 
European Theoretical Spectroscopy Facility (INFRA-2007-211956).
\end{acknowledgments}
\newpage
\appendix
\section{Fano resonances}
In this section we derive expressions for the Fano parameters\cite{fano} of the 
autoionizing resonances of single-particle character. The derivation 
is based on adiabatic linear response TDDFT.  

In the zeroth-order approximation, given here by the non-interacting 
KS system, there is no coupling between single-particle transitions. 
Different excitation channels can therefore be treated independently. 
When interactions are present and two channels overlap, one with 
discrete levels and the other with continuum levels, the 
coupling forms resonance structures having so-called Fano profiles. 

Starting from the assumption of having only one discrete transition 
superimposed on only one continuum, the KS response function can be written as
\be
\chi_s^R(\vr,\vr',z)=\frac{\tilde{f}_0(\vr)\tilde{f}_0(\vr')}{z^2-\ve^2_0}+\int^\infty_{I}\!\! 
d\ve\,\frac{\tilde{f}_\ve(\vr)\tilde{f}_\ve(\vr')}{z^2-\ve^2}.
\ee 
Choosing the retarded response function which is analytic in the upper 
half of the complex plane we have $z=\w+i0^+$. 
The quantities $f_0$ and  $\ve_0$ are the excitation function and the 
excitation energy of the discrete transition respectively. 
Notice that we have defined $\tilde{f}_0(\vr)=2\sqrt{\ve_0}f_0$ in order to 
make the formulas as light as possible. The ionization potential 
$I$ is naturally smaller than $\ve_0$. As an example of this model we can 
consider the $1s\to2p$ transition of Be where the other $1s\to np$ 
transitions are neglected and the continuum is given by $2s\to\ve p$. 

Within TDDFT the fully interacting $\chi^R$ can be obtained from Eq. 
(\ref{m2}):
\bea
\!\!\!\!\chi^R(\vr,\vr',z)&\!=\!&\tilde{f}_0(\vr)[z^2\mathbf{I}-\mathbf{V}]_{00}^{-1}\tilde{f}_0(\vr')\nn\\
&&\!+\!\int\!\! 
d\ve\,\tilde{f}_\ve(\vr)[z^2\mathbf{I}-\mathbf{V}]_{\ve 
0}^{-1}\tilde{f}_0(\vr')\nn\\
&&\!+\!\int\!\! 
d\ve'\,\tilde{f}_0(\vr)[z^2\mathbf{I}-\mathbf{V}]_{0\ve' 
}^{-1}\tilde{f}_{\ve'}(\vr')\nn\\
&&\!+\int\!\!d\ve d\ve'\, 
\tilde{f}_\ve(\vr)[z^2\mathbf{I}-\mathbf{V}]_{\ve\ve'}^{-1}\tilde{f}_{\ve'}(\vr').
\label{ovan}
\eea
The matrix $\mathbf{V}$ has a submatrix containing elements involving 
only excitations to the continuum. This matrix can be diagonalized initially 
by a linear transformation of all the continuum excitation functions $\tilde{g}_{\ve}=\int 
d\ve'U_{\ve\ve'}\tilde{f}_{\ve'}$. If the kernel is frequency independent (adiabatic) 
this transformation is unitary. The matrix $z^2\mathbf{I}-\mathbf{V}$ then takes the form
\be
\left(%
\begin{array}{cccc}
z^2-V_{00} & V_{0\ve'} & \ldots \\
V_{\ve0} & (z^2-\ve^2)\delta_{\ve\ve'}&\\
\vdots && \ddots 
\end{array}%
\right),
\label{fanom}
\ee
where $V_{00}=\tilde{\ve}_0^2=\ve_0^2+v_{00}$, and 
$V_{\ve0}=V_{0\ve}=-v_{0\ve}$.
From Sec.~\ref{lre}, 
$v_{00}=\bra\tilde{f}_0|v+f^A_{\xc}|\tilde{f}_0\ket$ and $v_{0\ve}=
\bra\tilde{f}_0|v+f^A_{\xc}|\tilde{g}_\ve\ket$, where the superscript $A$ signifies an
adiabatic approximation to $f_\xc$.

Analytic expressions for the elements of the inverse of  
$z^2\mathbf{I}-\mathbf{V}$ can now easily be obtained.
Defining the complex function
\be
B(\vr,z)=\int_I^{\infty} d\ve\frac{ 
v_{0\ve}\tilde{g}_\ve(\vr)}{z^2-\ve^2},
\ee
we can write $\chi^R$ as
\bea
\chi^R(\vr,\vr',z)&\!=\!&\frac{(\tilde{f}_0(\vr)+B(\vr,z))(\tilde{f}_0(\vr')+B(\vr',z))}{z^2-\tilde{\ve}_0^2-F(z)}\nn\\
&&+\chi^c_{s}(\vr,\vr',z),
\eea
where $\chi^c_{s}=\int^\infty_{I}\!\! 
d\ve\,\frac{\tilde{g}_\ve(\vr)\tilde{g}_\ve(\vr')}{z^2-\ve^2}$, i.e., 
the non-resonant 
background continuum and
\be
F(z)=\int_I^{\infty} d\ve\frac{v_{0\ve}^2}{z^2-\ve^2}.
\ee 

From now on, let us assume that we have integrated $\chi^R$ with some 
perturbing time-dependent potential and thereby removed the dependence on 
$\vr$ and $\vr'$. After some manipulations we can extract the real and imaginary parts 
of $\chi^R$. 
If  $\w> 0$ we have
\bea
{\rm Re}\,F\!&=&\!\mathcal{P}\int d\ve 
\frac{v_{0\ve}^2}{\w^2-\ve^2},\,\,\,\,{\rm Im}\,F=-\pi \frac{v_{0\w}^2}{2\w}\nn
\eea
Note that the real and imaginary parts of $F$ depend on $\w$. 
For $\w> 0$ we also have that 
\bea
{\rm Re}\,B\!&=&\!\mathcal{P}\int d\ve \frac{\tilde{f}_\ve 
v_{0\ve}}{z^2-\ve^2},\,\,\,\,{\rm Im}\,B=-\pi \frac{\tilde{g}_\w v_{0\w}}{2\w}\nn
\eea
If we define
\bea
\Gamma&=&-2{\rm 
Im}\,F,\,\,\,\,\epsilon=\frac{(\w^2-\tilde{\ve}_0^2-{\rm 
Re}\,F)}{\Gamma/2},\nn\\
q&=&-\frac{\tilde{f}_0+{\rm Re}\,B}{{\rm Im}\,B}
\label{parameters}
\eea
we can write
\begin{subequations}
\bea
{\rm Re}\,\chi^R(\w)&=&\frac{\epsilon(1-q^2)+2q}{\epsilon^2+1}{\rm 
Im}\,\chi_s^c+{\rm Re}\,\chi_s^c
\label{repart}\\
{\rm Im}\,\chi^R(\w)&=&\frac{(\epsilon+q)^2}{\epsilon^2+1}{\rm 
Im}\,\chi_s^c.
\label{impart}
\eea
\end{subequations}
\begin{subequations}
The full $\chi^R$ can then be written as
\bea
\chi^R(\w)&\stackrel{\w>0}{=}&\frac{(i\epsilon-q^2+2iq)}{(\epsilon+i)}{\rm 
Im}\,\chi_s^c+{\rm Re}\,\chi_s^c.
\eea
If $\w<0$ the first term in Eq. (\ref{repart}) 
acquires a minus sign and the full $\chi^R$ becomes
\bea
\chi^R(\w)&\stackrel{\w<0}{=}&\frac{(i\epsilon+q^2+2iq)}{(\epsilon-i)}{\rm 
Im}\,\chi_s^c+{\rm Re}\,\chi_s^c.
\eea
\end{subequations}
These equations show that $\chi^R$ has the correct analytic 
structure. It is analytic in the upper half of the 
complex plane and has a symmetric real part and an antisymmetric 
imaginary part. 
If we redefine  $F\to F/(\w+\tilde{\ve}_0)$ such that ${\rm 
Re}\,F\to{\rm Re}\,F/(\w+\tilde{\ve}_0)$ and 
$\Gamma\to\Gamma/(\w+\tilde{\ve}_0)$ the poles of $\chi^R$ are 
located at $\tilde{\ve}_0+{\rm Re}\,F\pm i\Gamma/2$, 
where the plus sign refers to $\w<0$. With a reasonably well behaved 
kernel the poles are in the lower half of the complex 
plane as they should. The real part corresponds to the position of 
the resonance and the imaginary part to the width. 

The formula in Eq. (\ref{impart}) gives the asymmetric resonance profile due 
to Fano but the parameters are here defined in terms of TDDFT 
quantities. 

With this derivation we have shown that Fano resonances for 
single-particle excitations occur already in RPA, i.e., 
with $f_{\xc}=0$, and how an adiabatic kernel modifies the 
parameters. (See also Ref. \onlinecite{maitrafano}) 

All parameters are frequency dependent, but  this is assumed to be 
weak and approximately constant over the resonance region.
Double and multiple-particle excitations are not  present in the adiabatic 
approximation since such states are absent in the KS response function. 
Maitra et. al.\cite{m1} have studied how $f_\xc$ should behave in order to 
describe double excitations. 

The above discussion has been limited to the case of one discrete state 
and one continuum. 
The formula can, however, easily be generalized to the case of several discrete 
states coupled to several continua. The derivation does not involve steps more 
complicated than those above and therefore we simply state the formulas and 
discuss the parameters. If we have $n$ discrete states and $m$ 
continua the response function can be generalized to ($\w>0$)
\bea
{\rm 
Re}\,\chi^R(\w)&=&\sum_{i=1}^n\frac{\epsilon_i(1-q_i^2)+2q_i}{\epsilon_i^2+1}\rho_i^2{\rm 
Im}\,\chi_s^c+{\rm Re}\,\chi_s^c\nn\\
{\rm 
Im}\,\chi^R(\w)&=&\sum_{i=1}^n\frac{(\epsilon_i+q_i)^2}{\epsilon_i^2+1}\rho_i^2{\rm 
Im}\,\chi_s^c+(1-\sum_{i=1}^n\rho_i^2){\rm Im}\,\chi_s^c
\label{chifull}
\eea
where $\chi_s^c=\sum_{k=1} ^m\chi_{s,k}^c$, i.e., a sum over the non-resonant 
background for 
every continuum channel. We see that we have different 
$q$-parameters for every discrete state as well as different widths. 
These are given by the new definitions of $F$ and $B$ which now become
$F_i=\sum_{k=1}^mF^k_i$ and $B_i=\sum_{k=1}^mB^k_i$.
The new parameter $\rho$ is defined as 
\be
\rho_i^2=-\frac{1}{\G_i/2}\frac{({\rm Im} B_i)^2}{{\rm Im}\chi_s^c}
\label{rho}
\ee
and gives the fraction of the continuum which mixes with the discrete 
state. If there is only one continuum $\rho=1$ for every discrete 
state and we have a zero on one side of the resonance.
In order to obtain Eq. (\ref{chifull}) for $\chi^R$ we have assumed that the matrix
\bea
M_{ij}&=&(z^2-\tilde{\ve}^2_i)\delta_{ij}-W_{ij},
\label{offdiag}
\eea
where
$$
W_{ij}=\sum_{k=1}^m\int_{I_k}^\infty d\ve_k\frac{v_{i\ve_k}v_{\ve_k 
j}}{z^2-\ve_k^2}
$$
is diagonal.  Physically, the matrix $W$ describes the interaction between the discrete 
states via the continua and the neglect of off-diagonal elements of this matrix has a very small effect
on the spectra as seen by comparing our parametrized spectra to the full ones in Sec.~\ref{consp}. 

The derivation above is based on adiabatic TDDFT. If the kernel is 
frequency dependent it also has an imaginary part which cannot be neglected. We notice that
with a general complex kernel it is not possible to derive the equations above using the same steps.


\begin{thebibliography}{50}
\bibitem[{fot()}]{fot1}
\bibinfo{note}{Notice that starting from the variational formulation based on
  the Klein functional leads to the necessity to invert the non-interacting
  response twice in order to construct the XC kernel. This is most likely not
  the case starting from a more sophisticated variational functional.}
\end{thebibliography}
\end{document}